\documentclass[letterpaper,11pt,aps,english,reprint,twocolumn,floatfix,nofootinbib,showpacs,superscriptaddress]{revtex4-2}

\usepackage[T1]{fontenc}
\usepackage[utf8]{inputenc}
\usepackage{babel}
\usepackage{amsmath}
\usepackage{calc}
\usepackage{amssymb}
\usepackage{latexsym}
\usepackage{soul}
\usepackage{color} 
\usepackage{colortbl}
\usepackage[table]{xcolor}
\usepackage{mathtools}
\usepackage{multirow}
\usepackage{microtype}
\usepackage{lmodern}

\usepackage{graphicx}
\usepackage{float}

\usepackage{color, xcolor}

\definecolor{myblue}{rgb}{0.00,0.50,2}
\definecolor{mymagenta}{rgb}{0.858, 0.188, 0.478}

\newcommand{\AddrUNAM}{ Instituto de F\'{\i}sica, Universidad Nacional Aut\'onoma de M\'exico, A.P. 20-364, Ciudad de M\'exico 01000, M\'exico}
\newcommand{\AddrAntioquia}{ Instituto de F\'{\i}sica, Universidad de Antioquia, A.A. 1226, Medell\'{\i}n, Colombia}
\newcommand{\AddrCinves}{Departamento de F\'isica, Centro de Investigaci\'on y de Estudios Avanzados del Instituto Polit\'ecnico Nacional\\
Apartado Postal 14-740, 07000 Ciudad de M\'exico, M\'exico}
\usepackage[unicode=true,pdfusetitle,
 bookmarks=true,bookmarksnumbered=false,bookmarksopen=false,
 breaklinks=false,pdfborder={0 0 1},backref=false,colorlinks=true]
 {hyperref}
\hypersetup{
 citecolor=myblue,linkcolor=mymagenta,urlcolor=mymagenta}

\begin{document}

\title{Two-zero textures for Dirac Neutrinos}

\author{Yessica Lenis}\email{yessica.lenis@udea.edu.co}\affiliation{\AddrAntioquia}\affiliation{\AddrUNAM}
\author{R. Martinez-Ramirez}\email{rolandomr@estudiantes.fisica.unam.mx}\affiliation{\AddrUNAM}
\author{Eduardo Peinado}\email{epeinado@fisica.unam.mx}\affiliation{\AddrUNAM}\affiliation{\AddrCinves}
\author{William A. Ponce}\email{william.ponce@udea.edu.co}\affiliation{\AddrAntioquia}

\begin{abstract} 
We review the two-zero mass matrix textures approach for Dirac neutrinos with the most recent global fit in the oscillation parameters. We found that three of the 15 possible textures are compatible with current experimental data, while the remaining two-zero textures are ruled out. Two textures are consistent with the neutrino masses' normal hierarchy and are CP-conserving. At the same time, the other one is compatible with both mass orderings and allows for CP violation. We also present the correlations between the oscillation parameters for the allowed two-zero textures. 
\end{abstract}


\maketitle

\section{\label{sec:introduccion} Introduction}

The Standard Model (SM) of particle physics is a successful theory that explains current experiments up to the TeV scale. Regardless of
this success, there are still open questions, such as the existence of dark matter~\cite{2009GReGr..41..207Z,1970ApJ...159..379R,Bertone:2004pz,Planck:2018vyg} where direct detection experiments have shown only null results~\cite{LUX:2016ggv,PandaX-II:2017hlx,XENON:2018voc,PICO:2019vsc}, baryon asymmetry in the universe~\cite{Sakharov:1967dj,Steigman:1976ev,Cohen:1997ac,Riotto:1999yt,Dine:2003ax}, and neutrino masses. From oscillation experiments~\cite{PhysRevLett.20.1205,Super-Kamiokande:1998kpq,SNO:2002tuh,KamLAND:2002uet,K2K:2006yov,MINOS:2006foh,KamLAND:2008dgz,T2K:2011ypd,DoubleChooz:2011ymz,DayaBay:2012fng,RENO:2012mkc,T2K:2014ghj}, we know neutrinos are massive particles. Nevertheless, these are sensitive only to the squared mass differences.  We still do not know the mass of any of the light neutrinos. Anyway, from cosmology, we know an upper limit for the sum of the three light neutrino masses, $\sum m_\nu$~\cite{Planck:2018vyg}, while from tritium beta decay, an upper limit for the effective mass of the electron neutrino~\cite{KATRIN:2019yun}. 

Another open question is whether neutrinos are Dirac particles, like the charged fermions or Majorana particles. If neutrinos are Majorana particles, there is a possibility to measure the neutrinoless double beta decay process~\cite{Schechter:1981bd}. This process has never been observed despite experimental effort~\cite{EXO-200:2019rkq,GERDA:2020xhi,CUORE:2021mvw,KamLAND-Zen:2022tow}, and the question remains unanswered. The alternative is to assume that massive neutrinos must be related to Dirac fields.

Besides the fact that no experiment has excluded so far the possibility of Dirac neutrino masses, there are several theoretical motivations to assume them, for example, the generation of baryon asymmetry via leptogenesis~\cite{Dick:1999je}, alternative approaches to the seesaw mechanism~\cite{Wang:2006jy}, from Peccei-Quinn symmetry~\cite{Peinado:2019mrn,delaVega:2020jcp} and the generation of radiative neutrino masses via quantum loops~\cite{mal,ma2,ma3,ma4,CentellesChulia:2019xky}. Also, in models derived from string theories, the Majorana masses are strongly suppressed by selection rules related to the underlying symmetries~\cite{langa}.

Considering physics beyond the standard model, there are several 
theoretical ways to obtain relationships between the fermion mixing angles and their physical masses, being the most popular ones the so-called horizontal symmetries (``top-down approach''), discrete~\cite{wilc1,babu,Aranda:2013gga,Bonilla:2017ekt} and continuos~\cite{wilc2,koca}, and texture zeros in the mass matrix in the flavor basis~\cite{fritz,rrr,iba} (``bottom-up approach''). Our analysis assumes the second approach without paying attention to the kind of physics implicit in the textures.

This article studies the phenomenology of Dirac neutrinos. The study is performed on the charged lepton diagonal mass basis, assuming the neutrino mass matrix to be Hermitian and with two vanishing elements.  

The number of vanishing elements in the neutrino mass matrix determines the type of texture. With one vanishing element, the Majorana neutrino case gives one zero equation and is compatible with the oscillation data. For two vanishing elements, only some cases are consistent with the experimental data~\cite{Frampton:2002yf}. 

If we consider the case of Dirac neutrinos whose mass matrix is Hermitian, there are 15 possible two-zero textures. Following the notation (for Majorana neutrinos) of Framptom {\it et. al.}~\cite{Frampton:2002yf} we can classify these 15 cases as:

\begin{equation}\label{A}
    A_1 : \begin{pmatrix}
    0 && 0 && \text{X} \\
    0 && \text{X} && \text{X} \\
    \text{X} && \text{X} && \text{X} 
    \end{pmatrix},
    \quad \quad A_2 : \begin{pmatrix}
    0 && \text{X} && 0 \\
    \text{X} && \text{X} && \text{X} \\
    0 && \text{X} && \text{X} 
    \end{pmatrix} ;
\end{equation}

\begin{equation}\label{B}
    B_1 : \begin{pmatrix}
    \text{X} && \text{X} && 0 \\
    \text{X} && 0 && \text{X} \\
    0 && \text{X} && \text{X}
    \end{pmatrix},
    \quad B_2 :
    \begin{pmatrix}
    \text{X} && 0 && \text{X} \\
    0 && \text{X} && \text{X} \\
    \text{X} && \text{X} && 0
    \end{pmatrix},
\end{equation}
\begin{equation}
    \quad B_3 : \begin{pmatrix}
    \text{X} && 0 && \text{X} \\
    0 && 0 && \text{X} \\
    \text{X} && \text{X} && \text{X} 
    \end{pmatrix},
      \quad B_4 : \begin{pmatrix}
    \text{X} && \text{X} && 0 \\
    \text{X} && \text{X} && \text{X} \\
    0 && \text{X} && 0 
    \end{pmatrix};
\end{equation}

\begin{equation}\label{C}
    C : \begin{pmatrix}
    \text{X} && \text{X} && \text{X} \\
    \text{X} && 0 && \text{X} \\
    \text{X} && \text{X} && 0
    \end{pmatrix};
\end{equation}

\begin{equation}\label{D}
        D_1 : \begin{pmatrix}
    \text{X} && \text{X} && \text{X} \\
    \text{X} && 0 && 0 \\
    \text{X} && 0 && \text{X} 
    \end{pmatrix},
    \quad \quad D_2 : \begin{pmatrix}
    \text{X} && \text{X} && \text{X} \\
    \text{X} && \text{X} && 0 \\
    \text{X} && 0 && 0
    \end{pmatrix} ;
\end{equation}

\begin{equation}\label{E}
    E_1 : \begin{pmatrix}
    0 && \text{X} && \text{X} \\
    \text{X} && 0 && \text{X} \\
    \text{X} && \text{X} && \text{X}
    \end{pmatrix},
    \quad E_2 :
    \begin{pmatrix}
    0 && \text{X} && \text{X} \\
    \text{X} && \text{X} && \text{X} \\
    \text{X} && \text{X} && 0
    \end{pmatrix},
\end{equation}
\begin{equation}
    E_3 : \begin{pmatrix}
    0 && \text{X} && \text{X} \\
    \text{X} && \text{X} && 0 \\
    \text{X} && 0 && \text{X} 
    \end{pmatrix}
    \text{ }\text{ and }\text{ }F_1 : \begin{pmatrix}
    \text{X} && 0 && 0 \\
    0 && \text{X} && \text{X} \\
    0 && \text{X} && \text{X}
    \end{pmatrix},
\end{equation}
\begin{equation}\label{F}
    F_2 :
    \begin{pmatrix}
    \text{X} && 0 && \text{X} \\
    0 && \text{X} && 0 \\
    \text{X} && 0 && \text{X}
    \end{pmatrix},
    \quad F_3 : \begin{pmatrix}
    \text{X} && \text{X} && 0 \\
    \text{X} && \text{X} && 0 \\
    0 && 0 && \text{X}
    \end{pmatrix}.
\end{equation}
The $\text{X}$ represents a non-vanishing entry in Eqs.~\eqref{A}-\eqref{F}. The zeros in the neutrino mass matrix imply correlations among the neutrino oscillation parameters, as discussed in the next section. We will show that the only textures compatible with current data are $A_1$, $A_2$, and $C$.

While numerous studies have explored texture zeros in Majorana neutrinos~\cite{Xing:2002ta,Fritzsch:2011qv,Meloni:2014yea,2016ChPhC..40c3102Z,Alcaide:2018vni,Singh:2019baq}, the Dirac neutrinos scenario has received less attention. The phenomenological implications of texture zeros in the neutrino mass matrix for the Dirac case in the charged leptons diagonal mass basis were studied previously, considering the implications of zeros in symmetric~\cite{Hagedorn:2005kz}, general mass matrices~\cite{Borgohain:2020csn} and Hermitian~\cite{Liu:2012axa,Borgohain:2020csn}. This work aims to relate the observables based on the analytical expressions derived from the vanishing elements. In a model-independent approach, we employ a straightforward numerical technique involving scanning within the $3\sigma$ region, obtaining correlations between some of the oscillation parameters. Our analysis reveals discrepancies with previous studies that have assumed a Hermitian neutrino mass matrix, highlighting the need for further exploration in this area.

The following sections provide an overview of our general approach and discuss some implications of the previous textures in normal ordering (NO) and inverted ordering (IO) of the neutrino mass spectrum. Finally, we present our numerical analysis and the correlations resulting from the numerical analysis for the favored textures. 

\section{\label{sec:Desarrollo} General Approach}
In the charged lepton diagonal mass basis, for a Hermitian  Dirac neutrinos mass matrix, $M_{\nu}$, the Pontecorvo, Maki, Nakagawa, Sakata (PMNS) matrix $U_{\text{PMNS}}$ diagonalizes it, in the following way
\begin{equation}\label{diag}
U_{\text{PMNS}}^\dagger \, M_{\nu} \, U_{\text{PMNS}} =  \text{Diag} ({m_{1}},{m_{2}},{m_{3}}).
\end{equation}
The standard parametrization of the PMNS mixing matrix makes use of the three mixing angles ($\theta_{12}, \theta_{13}, \theta_{23}$) and the Dirac phase ($\delta$), which is responsible for CP violation~\cite{Workman:2022ynf}
$ \\ $
\begin{widetext}
\begin{equation}
U_{\text{PMNS}} \equiv U = \begin{pmatrix}
U_{e1} && U_{e2} && U_{e3} \\
U_{\mu 1} && U_{\mu 2} && U_{\mu 3} \\
U_{\tau 1} && U_{\tau 2} && U_{\tau 3} 
\end{pmatrix} \\\
= \begin{pmatrix*}[r]
1 && 0 && 0 \\
0 && c_{23} && s_{23} \\
0 && -s_{23} && c_{23} 
\end{pmatrix*}
\begin{pmatrix*}[c]
c_{13} && 0 && s_{13} e^{-i \delta} \\
0 && 1 && 0 \\
-s_{13} e^{i \delta} && 0 && c_{13} 
\end{pmatrix*}
  \begin{pmatrix*}[r]
c_{12} && s_{12} && 0 \\
-s_{12} && c_{12} && 0 \\
0 && 0 && 1
\end{pmatrix*},
\end{equation}
\end{widetext}
where $c_{jk}=\cos\theta_{jk}$ and $s_{jk}=\sin\theta_{jk}$ respectively. 

From Eq.~\eqref{diag}, we can write the neutrino mass matrix in terms of the three neutrino masses and the mixing parameters as 
\begin{equation}\label{sourceofeqs}
M_{\nu}=U_{\text{PMNS}} \, \text{Diag} ({m_{1}},{m_{2}},{m_{3}}) \, U_{\text{PMNS}}^\dagger.
\end{equation}
In this way, the right-hand side of Eq.~\eqref{sourceofeqs} depends only on the neutrino mixing angles, the CP-violating phase, and the neutrino masses, forcing the neutrino mass matrix to reproduce the neutrino mixing compatible with the observed values. From here, it is clear that in Eq.~\eqref{sourceofeqs}, a zero on the left-hand side will imply an equation relating neutrino masses with the CP-phase and the neutrino mixing angles.

The Dirac mass matrices in the SM are arbitrary $3\times 3$ matrices. However, due to the polar theorem of linear algebra, it can be written as the product of a Hermitian matrix times a unitary matrix~\cite{Branco:2007nn}. Moreover, this latter unitary matrix can be absorbed in a redefinition of the right-handed fields since there is no limitation for the form of these fields in the SM (something forbidden in a left-right symmetric extension of the model). In this way, we can take the mass matrix for Dirac neutrinos to be Hermitian without losing generality.

A general $3\times 3$ Hermitian matrix has six real parameters and three phases. By a weak basis transformation, we can reduce the number of phases to one, ending up with a Hermitian matrix with six real parameters and one phase. It is enough to accommodate the seven physical parameters: the three neutrino masses, the three mixing angles, and the Dirac CP-violating phase.

Forcing a vanishing element in the neutrino mass matrix reduces the number of free parameters. One texture zero establishes a relationship between physical parameters, {\it i.e.}, mixing angles, and neutrino masses. With two vanishing elements, there are two relations. It is important to note that these relationships may or may not agree with the current experimental oscillation data. We study the phenomenology of each texture to see this compatibility.

\subsection*{Analysis}

We obtain a system of two equations (the corresponding to the vanishing entries of $M_{\nu}$), {\it i.e.}, $(M_{\nu})_{ij} \equiv M_{ij} = 0$ and $(M_{\nu})_{kl} \equiv M_{kl} = 0$ with $ij \neq kl$.  These equations allow us to write two neutrino masses as functions of the other neutrino mass (we choose the lightest) and the neutrino oscillation parameters. For NO (IO), the equations are of the form~\cite{Liu:2012axa}:

\begin{equation} \label{Ecuaciones}
\begin{aligned}
\eta m_{1}U_{i1}U_{1j}^* + \kappa m_{2}U_{i2}U_{2j}^* + m_{3}
U_{i3}U_{3j}^* &= 0,\\
\eta m_{1}U_{k1}U_{1l}^* + \kappa m_{2}U_{k2}U_{2l}^* + m_{3}
U_{k3}U_{3l}^* &= 0,
\end{aligned}
\end{equation}
where  $\eta$ and $\kappa$ are the relative $m_{1}$ and $m_{2}$ signs respect to $m_3$.
The value of $\delta$ in Eqs.~\eqref{Ecuaciones} is limited by the realness of the neutrino masses. The result will indicate whether or not CP is conserved:

\begin{itemize}
    \item \textbf{CP violation:} The neutrino mass matrices with both zeros in the diagonal allow CP violation. In such a case,  $i=j$ and $k=l$, and the Eqs.~\eqref{Ecuaciones} are both real, no matter the value of the $\delta$ CP phase. The textures that allow CP violation are $C$, $E_1$, and $E_2$ in Eqs.~\eqref{C} and~\eqref{E}.

    \item \textbf{CP conservation:} The remaining textures lead to CP conservation. In Eq.~\eqref{Ecuaciones}, the realness of the neutrino masses forces $\delta$ to take the $0$ or $\pi$ values.
    In Appendix~\ref{Appendix}, we provide an alternative proof of CP conservation when there is a zero off-diagonal matrix element.

\end{itemize}

We use the squared mass differences to write two masses in terms of the lightest neutrino mass:
\begin{equation}\label{Cereal}
\begin{aligned}
\Delta m_{21}^2 &=  m_{2}^2 - m_{1}^2, \\ 
\Delta m_{31}^2 &=  m_{3}^2 - m_{1}^2.
\end{aligned}
\end{equation}
We then give random values to the square mass differences in the $3\sigma$ range from the neutrino oscillations global fit~\cite{deSalas:2020pgw,Esteban:2020cvm}. After this, we assign numerical values to the lightest neutrino masses. The masses take real values since the mass matrix is Hermitian. With this in mind, we have found that three textures are compatible with the current neutrino oscillation data. Two textures, $A_1$ and $A_2$, are CP conserving (those textures have an off-diagonal vanishing matrix element) and are compatible only for normal ordering. The third one is texture C. This texture is compatible with CP violation for both neutrino mass hierarchies.

The following section presents our results for textures A1, A2, and C.

\section{Results and Conclusions}

Only three of these textures are compatible with current experimental data. Textures A1 and A2 are compatible with the normal ordering of the neutrino masses, while texture C is compatible with both hierarchies. These textures show clear correlations between some observables depending on the particular texture. We present the results of the neutrino phenomenology concerning such correlations in Figs.~\ref{ResultsA1}-\ref{ResultsCNO}, using the global fit from~\cite{deSalas:2020pgw}, given in Table~\ref{best_fit}.
\begin{table}[H]
\begin{center}
\begin{tabular}{| c || c | c |}
  \hline 
   \hspace{0.1cm}  Parameters    \hspace{0.1cm}       &    Best Fit $\pm 1 \sigma$    &    \hspace{0.2cm}   $3 \sigma$  Range      \hspace{0.2cm}  \\
\hline \hline
& & \\
$\Delta m_{21}^2: [10^{-5} \rm eV^2] $  &   $7.50^{+0.22}_{-0.20}$  &   $6.94 - 8.14$  \\	
 & & \\
$| \Delta m_{31}^2|: [10^{-3} \rm eV^2] $ (NO)  &   $2.55^{+0.02}_{-0.03}$  &   $2.47 - 2.63$  \\
& & \\
$| \Delta m_{31}^2|: [10^{-3} \rm eV^2] $ (IO)  &   $2.45^{+0.02}_{-0.03}$  &   $2.37 - 2.53$  \\
& & \\
$ \sin^2 \theta_{12} / 10^{-1}$  &   $3.18 \pm 0.16 $  &   $2.71 - 3.69$  \\
& & \\
$ \sin^2 \theta_{23} / 10^{-1}$ (NO)  &   $5.74 \pm 0.14 $  &   $4.34 - 6.10$  \\
& & \\
$ \sin^2 \theta_{23} / 10^{-1}$ (IO)  &   $5.78^{+0.10}_{-0.17} $  &   $4.33 - 6.08$  \\
& & \\
$ \sin^2 \theta_{13} / 10^{-2}$ (NO)  &   $2.200^{+0.069}_{-0.062} $  &   $2.000 - 2.405$  \\
& & \\
$ \sin^2 \theta_{13} / 10^{-2}$ (IO)  &   $2.225^{+0.064}_{-0.070} $  &   $2.018 - 2.424$  \\
& & \\
$ \delta_{CP} / \pi$ (NO)  &   $1.08^{+0.13}_{-0.12} $  &   $0.71 - 1.99$  \\
& & \\
$ \delta_{CP} / \pi$ (IO)  &   $1.58^{+0.15}_{-0.16} $  &   $1.11 - 1.96$  \\
& & \\
    \hline
  \end{tabular}
\end{center}
\caption[Global fit for neutrino oscillation parameters]{Global fit for neutrino oscillation parameters, given by \cite{deSalas:2020pgw}. }
 \label{best_fit}
\end{table} 
We contrast the neutrino masses with the upper limits from KATRIN collaboration~\cite{KATRIN:2019yun} ($m_{\nu} < 1.1 \text{ eV}$, with $m_{\nu}$ being the absolute mass scale for neutrinos), PLANCK collaboration~\cite{Planck:2018vyg} ($\sum m_{\nu} < 0.12  \text{ eV}$) and the most constraining bound up to date from cosmology of $ \sum m_{\nu} < 0.09 \text{ eV}$~\cite{DiValentino:2021hoh}.

\subsection*{Texture A1}

The resulting equations are:
\begin{align}
    m_2 &= m_1 \dfrac{c_{12} \left( e^{i \delta } c_{23} s_{12} s_{13} + c_{12} s_{23} \right)}{s_{12} \left( e^{i \delta } c_{12} c_{23} s_{13} - s_{12} s_{23} \right)}, \\
    m_3 &= m_1 \dfrac{ e^{i \delta } c_{12} \, c_{13}^2 \, c_{23} }{s_{13} \left( - e^{i \delta } c_{12} c_{23} s_{13} + s_{12} s_{23} \right)},
\end{align} 
where, to ensure real neutrino masses, the $\delta$ CP phase has to be $0$ or $\pi$. We found that this texture is only compatible with normal ordering. 

\begin{figure}[H]
 \includegraphics[width=0.92\linewidth,height=0.8\textheight]{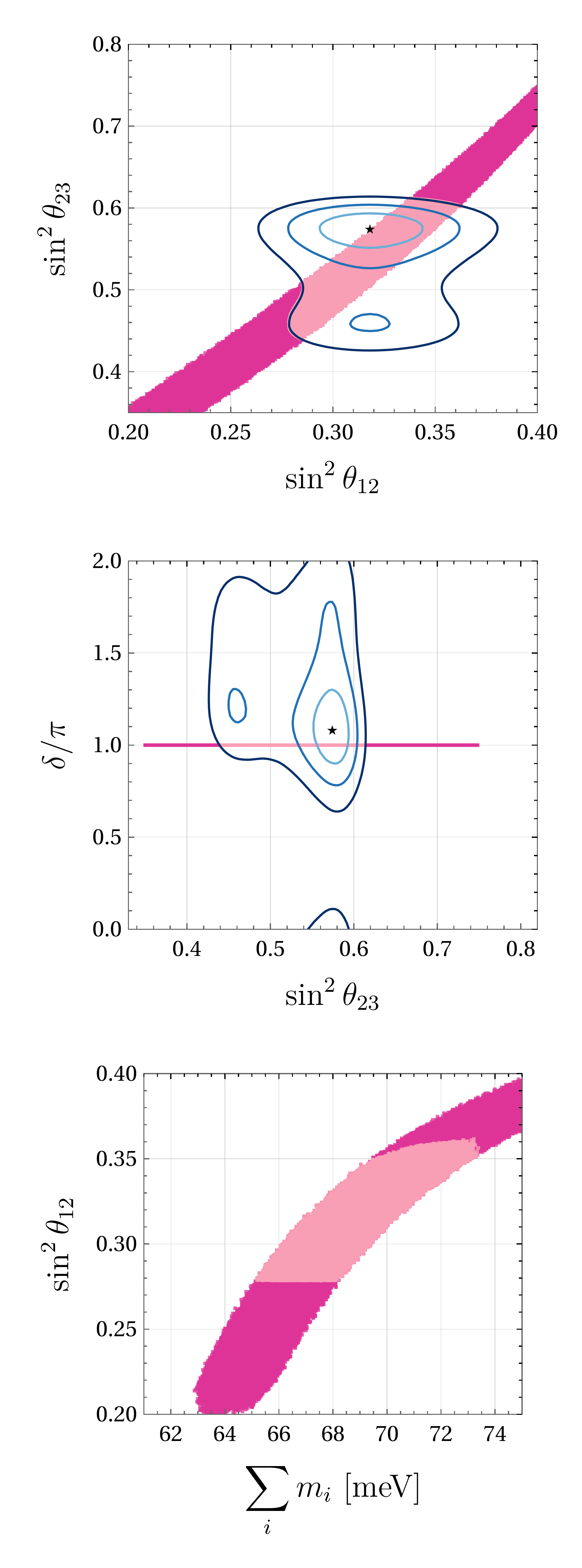}   
\caption{Correlations for texture A1 ($M_{ee} = M_{e \mu} = 0$) in NO. From top to bottom: $\sin^2 \theta_{12}$ vs. $\sin^2 \theta_{23}$,  $\sin^2 \theta_{23}$ vs. $\delta$ and $\sum m_{\nu}$ vs. $\sin^2 \theta_{12}$. The upper and middle plots have blue contours corresponding to $3\sigma$, $2\sigma$, and $1\sigma$ along with the best-fit point from \cite{deSalas:2020pgw}. Points in lighter pink represent solutions inside the contours also consistent with the bounds from \cite{KATRIN:2019yun}, \cite{Planck:2018vyg} and \cite{DiValentino:2021hoh}.  }
\label{ResultsA1}
\end{figure}

The upper panel in Fig.~\ref{ResultsA1} shows a correlation between the solar and atmospheric mixing angles, showing a wide overlap region with the global fit contours. As the parameter $\sin^2{\theta_{12}}$ increases, there is a corresponding increase in $\sin^2{\theta_{23}}$. In the middle panel of Fig.~\ref{ResultsA1}, $\delta = \pi$ indicates CP conservation. The lower panel in Fig.~\ref{ResultsA1} shows the correlation between $\sum m_i$ and the solar mixing angle. For this texture, the mass values are consistent with all the upper limits.  

\subsection*{Texture A2}

\begin{figure}[H]
\includegraphics[width=0.92\linewidth,height=0.8\textheight]{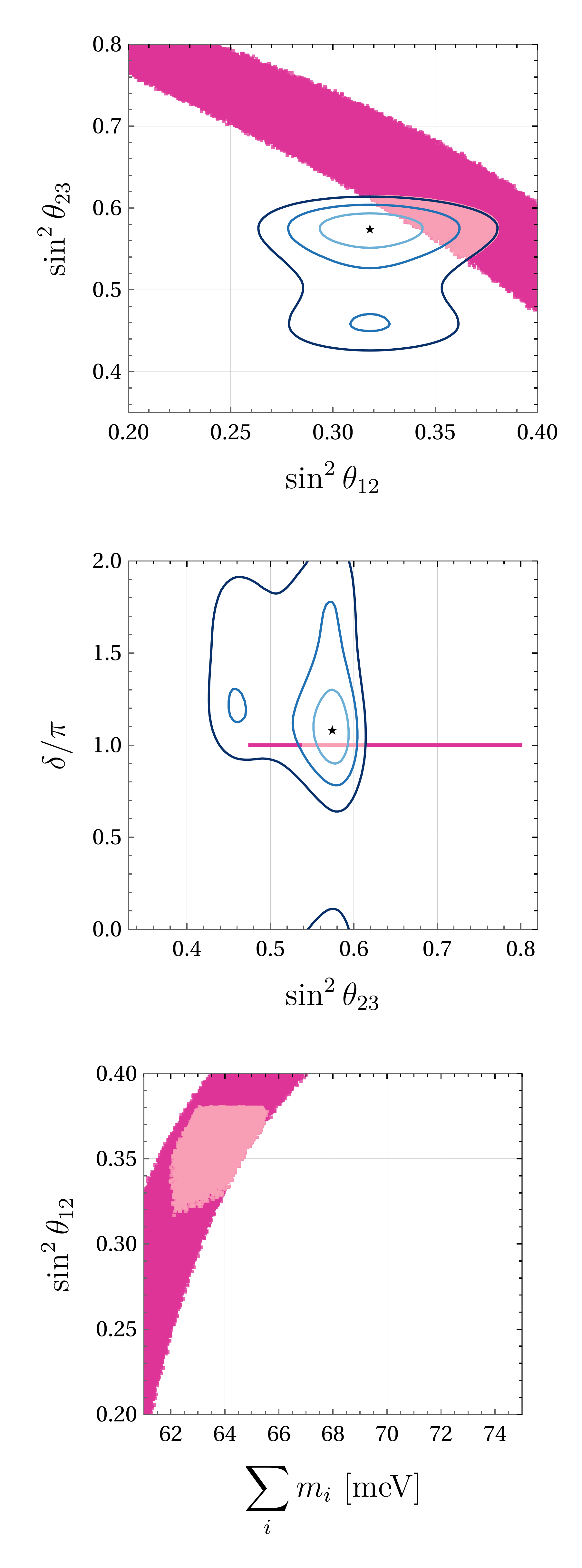}  
\caption{Correlations for texture A2 ($M_{ee} = M_{e \tau} = 0$) in NO. From top to bottom: $\sin^2 \theta_{12}$ vs. $\sin^2 \theta_{23}$, $\sin^2 \theta_{23}$ vs. $\delta$ and $\sum m_{\nu}$ vs. $\sin^2 \theta_{12}$. Same color code as Fig.~\ref{ResultsA1}.}
\label{ResultsA2}
\end{figure}

In this case, the resulting equations are:
\begin{align}
    m_2 &= m_1 \dfrac{c_{12} \left( - c_{12 } c_{23} + e^{i \delta } s_{12} s_{13} s_{23} \right)}{s_{12} \left( c_{23} s_{12} + e^{i \delta } c_{12} s_{13} s_{23} \right)}, \\
    m_3 &= - m_1 \dfrac{e^{i \delta } c_{12} \, c_{13}^2 \, s_{23} }{s_{13} \left( c_{23} s_{12} + e^{i \delta } c_{12} s_{13} s_{23} \right)}.
\end{align} 
Again, $\delta$ has to be $0$ or $\pi$ to obtain real neutrino masses. As in texture A1, this texture is only compatible with normal ordering. The upper panel of Fig.~\ref{ResultsA2} shows the correlation between the solar and atmospheric mixing angles. 

In contrast to texture A1, a decreasing in $\sin^2{\theta_{23}}$ implies an increasing in $\sin^2{\theta_{12}}$. The middle panel in Fig.~\ref{ResultsA2} indicates CP conservation. The overlap with the contours is also limited to greater values for the angles in this case. Lower panel of Fig.~\ref{ResultsA2} presents the neutrino mass correlation consistent with the upper limits. 

\subsection*{Texture C}

In the case of texture C, the equations are the following:
\begin{widetext}
\begin{align}
    m_1 &= m_3 \dfrac{c_{12} \, c_{13}^2 \left[ 2 \cos{\delta} \, c_{23} \, s_{12} \, s_{13} \, s_{23} + c_{12} \left( s_{23}^2 - c_{23}^2 \right) \right]}{s_{13} \left[ -2  \cos{\delta} \, c_{12} \,c_{23} \, s_{12} \left( 1 + s_{13}^2 \right) s_{23}  + \left( c_{12}^4 - s_{12}^4 \right) s_{13} \left( c_{23} - s_{23} \right) \left( c_{23} + s_{23} \right) \right]}, \\
    m_2 &= m_3 \dfrac{s_{12} \, c_{13}^2 \left[ - 2 \cos{\delta} \, c_{12} \, c_{23} \, s_{13} \, s_{23} + s_{12} \left( s_{23}^2 - c_{23}^2 \right)  \right] }{s_{13} \left[ 2  \cos{\delta}  \, c_{12} \,c_{23} \, s_{12} \left( 1 + s_{13}^2 \right) s_{23} - \left( c_{12}^4 - s_{12}^4 \right) s_{13} \left( c_{23} - s_{23} \right) \left( c_{23} + s_{23} \right) \right]}.
\end{align}
\end{widetext}
Texture C is compatible with both hierarchies, as we discuss below.
$ \\ $
$ \\ $
\textit{Inverse ordering:}

\begin{figure}
\centering
    \includegraphics[width=0.97\linewidth]{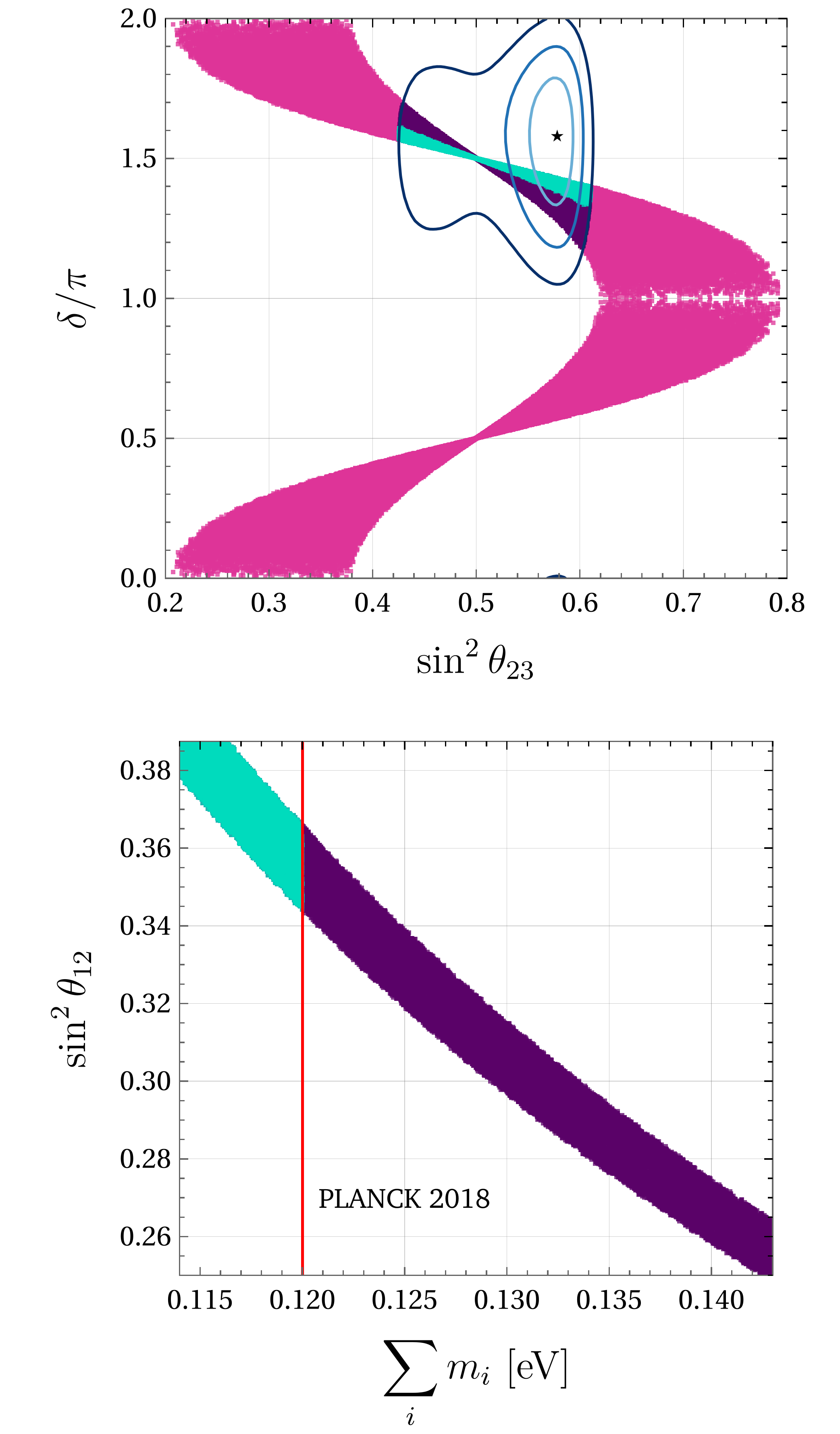}   
    \caption{Correlations for texture C ($M_{\mu \mu} = M_{\tau \tau} = 0$) in IO. In the upper plot, $\sin^2 \theta_{23}$ vs. $\delta$. In the lower plot, $\sum m_{\nu}$ vs. $\sin^2 \theta_{12}$. The $3\sigma$, $2\sigma$, and $1\sigma$ contours and the best-fit point are shown in the upper plot. The red vertical line represents the upper constraint given by the PLANCK collaboration \cite{Planck:2018vyg}. Both plots show points inside the contours that fulfill this condition in cyan color. Points in purple represent solutions inside the contours, also consistent with the KATRIN limit but not with PLANCK.}
    \label{ResultsCIO}
\end{figure}

The correlation between the atmospheric angle and the CP phase is displayed in the upper panel of Fig.~\ref{ResultsCIO} for the inverse ordering. 
Solutions inside the contours give a value for the CP phase $\delta \approx 3 \pi / 2$, leading to CP violation. The lower panel in Fig.~\ref{ResultsCIO} shows the negative correlation of the solar mixing angle with $\sum m_i$. Mass values printed in purple are consistent only with the bound from KATRIN, while cyan points are also consistent with the PLANCK bound. However, all points are ruled out by the limit imposed by the Cosmological bound from~\cite{DiValentino:2021hoh}.
$ \\ $
$ \\ $
\textit{Normal ordering:}

\begin{figure}
\centering
    \includegraphics[width=0.97\linewidth]{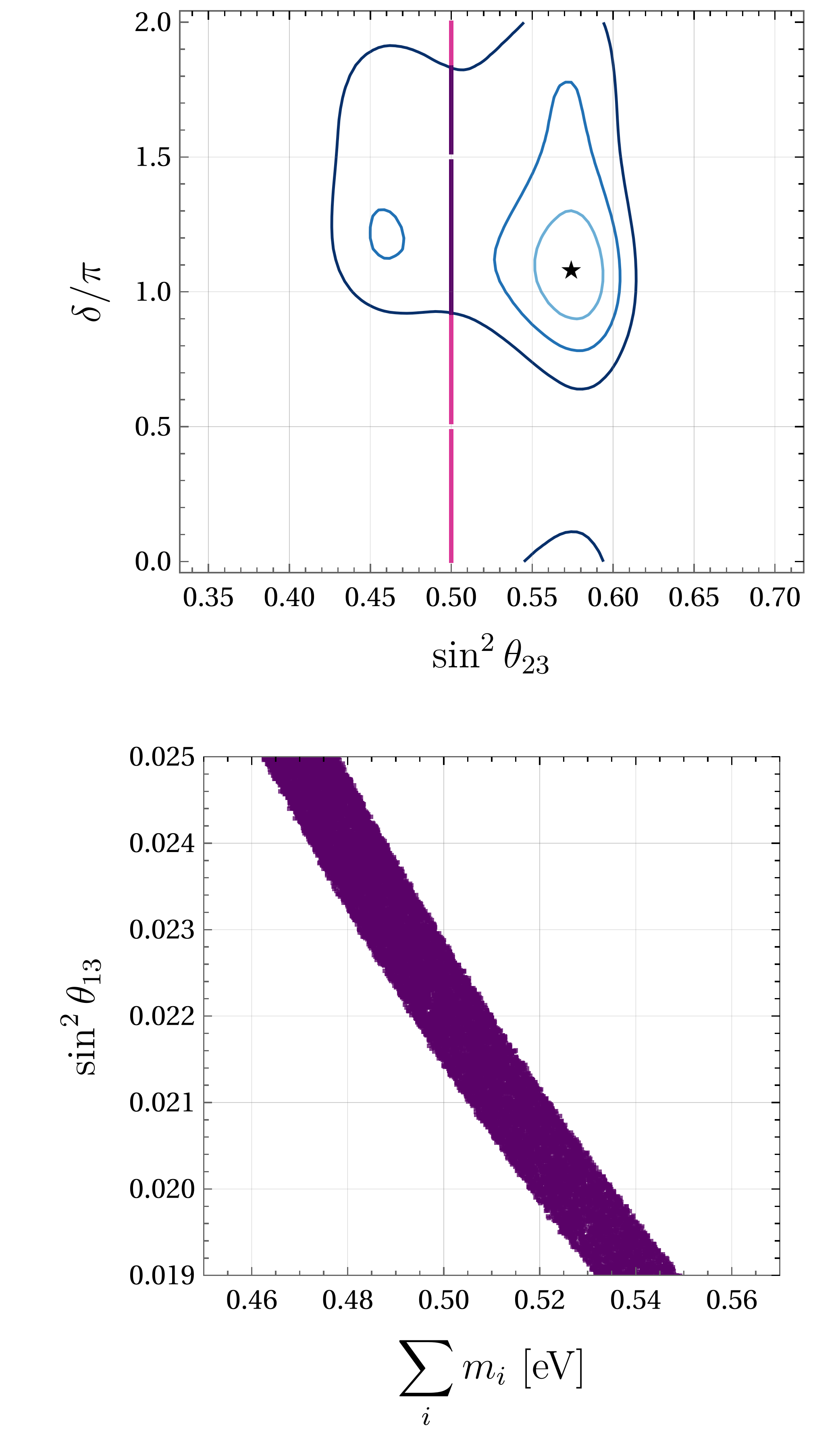} 
    \caption{Correlation for texture C ($M_{\mu \mu} = M_{\tau \tau} = 0$) in NO. In the upper plot, $\sin^2 \theta_{23}$ vs $\delta$. In the lower plot, $\sum m_{\nu}$ vs. $\sin^2 \theta_{13}$.  Same color code as Fig.~\ref{ResultsCIO}.}
    \label{ResultsCNO}
\end{figure}

The upper panel in Fig.~\ref{ResultsCNO} shows the correlation between the atmospheric angle and the CP phase for normal ordering.
Compared to inverse ordering, there is a higher range for CP-phase values inside the contours. Also, the maximal $\theta_{23} \approx \pi / 4$ is preferred, implying quasi-degeneration between neutrino masses. The correlation between the reactor mixing angle and $\sum m_i$ is displayed in the lower panel of Fig.~\ref{ResultsCNO}. The mass values are greater than those in inverse ordering, consistent only with the KATRIN limit. 

\subsection*{Ruled out Textures}

\begin{figure}
\centering
 \includegraphics[width=0.96\linewidth,height=0.31\textheight]{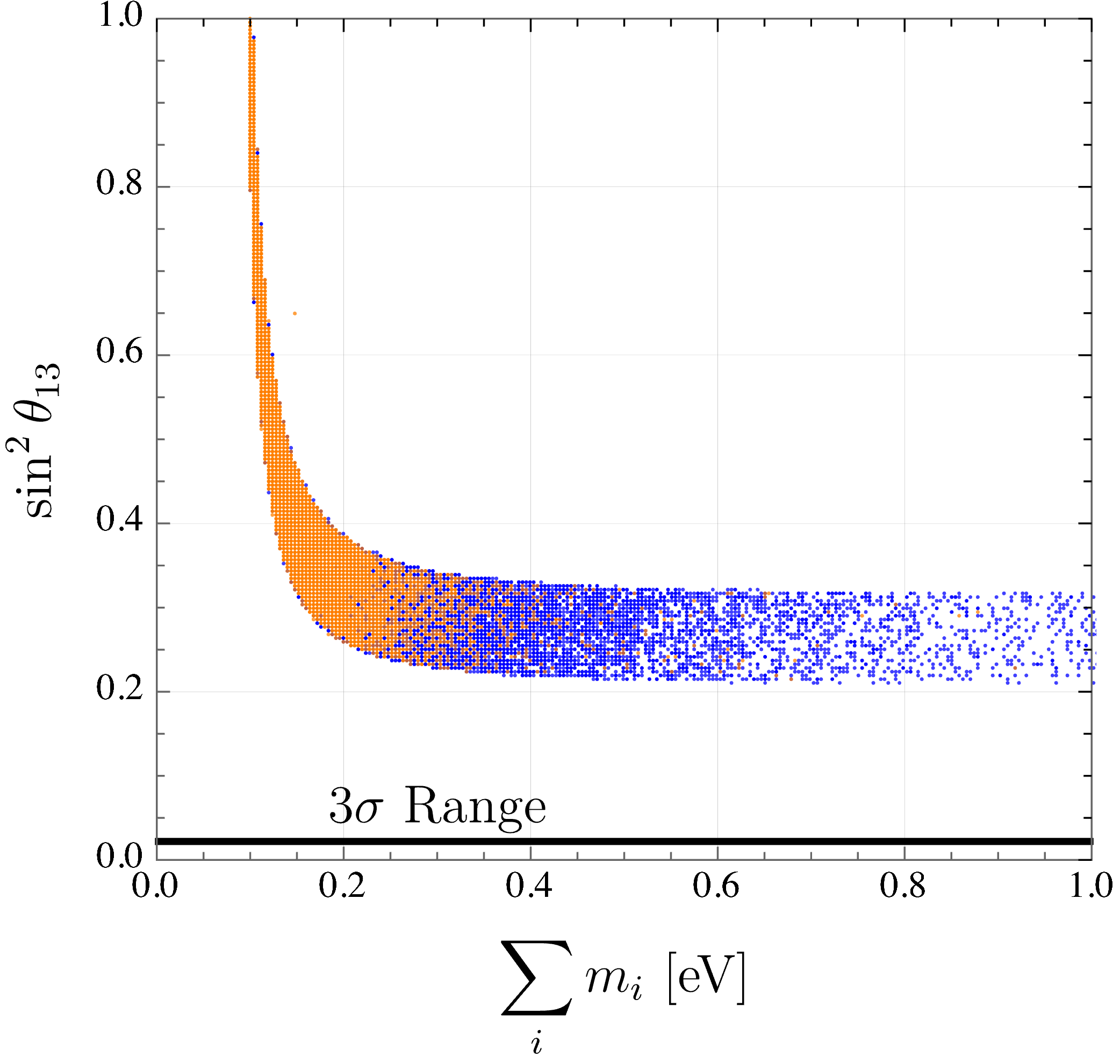}   
\caption{Correlation of $\sin^2 \theta_{13}$ vs. $\sum m_{\nu}$ for texture A1 ($M_{ee} = M_{e \mu} = 0$) in blue, and texture A2 ($M_{ee} = M_{e \tau} = 0$) in orange, for IO. The $3\sigma$ range for $\sin^2 \theta_{13}$ is displayed in black solid lines. }
\label{A1_A2}
\end{figure}

Textures A1 and A2 in inverse ordering are not allowed by current experimental data. This can be seen in Fig.~\ref{A1_A2},
where a correlation between the sum of masses and the reactor mixing angle is displayed for both textures. In black, the corresponding $3\sigma$ range from the global fit~\cite{deSalas:2020pgw} is shown without overlap with the points allowed for the rest of the mixing parameters. In textures B, for values of $\sin^2 \theta_{23}$ and $\sin^2 \theta_{13}$ within the $3\sigma$ range, the solutions for $\sin^2 \theta_{12}$ are either $\sim 0$ or $\sim 1$ for both hierarchies, contradicting experimental evidence. For textures D in the normal hierarchy, the resulting values for $\sin^2 \theta_{13}$ were $\gtrsim 0.3$ when scanning the other mixing angles in the $3\sigma$ range, therefore excluded. In inverse hierarchy, texture D1 led to $\sin^2 \theta_{23} \gtrsim 0.7$, while texture D2 led to $\sin^2 \theta_{23} \lesssim 0.25$, out of the global fit $3\sigma$ range. A similar pattern is observed with textures E1 and E2 for both hierarchies and E3 in normal ordering, where either $\sin^2 \theta_{13}$ or $\sin^2 \theta_{23}$ are found to be excluded by the global fit. In E3, no available parameter space was found for inverse ordering. Finally, texture F implies exactly degenerate masses.

$ \\ $
$ \\ $

Our results, as those in~\cite{Liu:2012axa}, found that textures A1, A2, and C are compatible with the current experimental neutrino oscillation data. However, we present the complete analysis for texture C in NO and obtain different correlations between the oscillation parameters for texture C in IO. The work in~\cite{Borgohain:2020csn} considers two-zero textures with Hermitian and non-Hermitian Dirac neutrino mass matrices. In the Hermitian case, they found that textures A1 and C are allowed, while A2 is ruled out, in contrast with the findings of our work.

In conclusion, the cases compatible with the experimental data are textures A1 and A2 for normal ordering and C for both hierarchies.  Textures A1 and A2, with normal ordering, predict CP conservation, while texture C predicts CP violation in both hierarchies. Texture C, with normal ordering, leads to quasi-degenerate neutrino masses.

Finally, the iterative use of weak basis transformations~\cite{Workman:2022ynf} led us to eliminate redundant phases in the general Hermitian neutrino mass matrix, leaving only one physical phase linked to possible CP violation in the lepton sector.

\section*{Acknowledgements}
This work is supported by the Mexican grants CONAHCYT CB-2017-2018/A1-S-13051 and DGAPA-PAPIIT IN111625. E.P. is grateful for funding from `C\'atedras Marcos Moshinsky' (Fundaci\'on Marcos Moshinsky) and for the support of PASPA-DGAPA, UNAM for a sabbatical leave. RMR thanks CONAHCYT for the funding of his MSc studies. YL thanks the Theoretical Physics Department of IFUNAM for the hospitality during her research stay.

\appendix 

\section{} \label{Appendix} 
In this appendix, we use the weak basis transformation to reduce the number of phases from three to one in a general $3\times 3$ Hermitian neutrino mass matrix. Furthermore, we discuss the reason for no CP violation when there is an off-diagonal vanishing element.

In the context of the SM extended with right-handed neutrinos and lepton number  conservation, the most general 
weak basis transformation that leaves the two $3\times 3$ lepton mass 
matrices Hermitian, and does not alter the physics implicit in the weak currents (does not alter the physical content in the PMNS  mixing matrix), is an 
arbitrary unitary transformation $U$ acting simultaneously in the charged lepton 
and in the neutrino mass matrices \cite{Branco:2007nn}. That is
\begin{equation}\label{2aa}
\begin{split}
M_\nu&\longrightarrow M_\nu^R=U M_\nu U^\dag,\\
M_l&\longrightarrow M_l^R=U M_l U^\dag.
\end{split}
\end{equation}

Now, when the mass matrices for the charged lepton sector are diagonal,  
we have that the most general Hermitian mass matrix $M_\nu$  for the neutral sector has six real parameters and three phases that we can use to explain seven physical parameters: three neutrino masses  $m_1, m_2$ and $m_3$, the three mixing angles $\theta_{12}, \theta_{13}$ and $\theta_{23}$,
and one CP violating phase $\delta$ in the PMNS mixing matrix. So, in principle, 
we have a redundant number of parameters (two more phases).

Contrary to the quark sector~\cite{ponce:2011qp,ponce:2013nsa}, we can not introduce texture zeros via weak 
basis transformations in the mass matrix $M_\nu$ since we already work in the charged lepton diagonal mass matrix. However, the ``weak basis transformations'' (WKB) can eliminate 
the redundant phases. To do this, let us write the neutrino mass matrix as:

\begin{equation}
M_{\nu}=\begin{pmatrix*}[l]\label{mngenp}
\vert m_{\nu_e\nu_e} \vert && \vert m_{\nu_e\nu_\mu} \vert e^{i\phi_{e\mu}} && \vert m_{\nu_e\nu_\tau} \vert e^{i\phi_{e\tau}} \\
\vert m_{\nu_e\nu_\mu} \vert e^{-i\phi_{e\mu}} && \vert m_{\nu_\mu\nu_\mu} \vert && \vert m_{\nu_\mu\nu_\tau} \vert e^{i\phi_{\mu\tau}} \\
\vert m_{\nu_e\nu_\tau} \vert e^{-i\phi_{e\tau}} && \vert m_{\nu_\mu\nu_\tau} \vert e^{-i\phi_{\mu\tau}} && \vert m_{\nu_\tau\nu_\tau} \vert
\end{pmatrix*};
\end{equation}

and let us do a weak basis transformation using the following diagonal unitary 
matrix:
\begin{equation}
\begin{aligned}
\displaystyle M_\phi&=\text{Diag}(e^{i\phi_1},1,e^{i\phi_2}), \\
M_\phi^\dagger&=\text{Diag}(e^{-i\phi_1},1,e^{-i\phi_2})=M_\phi^{-1},
\end{aligned}
\end{equation}
which does not change the diagonal charged lepton mass matrix. 
Afther this, the matrix \ref{mngenp} gets the following form:
\begin{widetext}
\begin{equation}\nonumber
M_\nu^\prime= \begin{pmatrix*}[l]
|m_{\nu_e\nu_e}| && |m_{\nu_e\nu_\mu}|e^{i(\phi_{e\mu}-\phi_1)} && |m_{\nu_e\nu_\tau}|e^{i(\phi_{e\tau}+\phi_2-\phi_1)} \\ 
|m_{\nu_e\nu_\mu}|e^{-i(\phi_{e\mu}-\phi_1)} && |m_{\nu_\mu\nu_\mu}| && |m_{\nu_\mu\nu_\tau}|e^{i(\phi_{\mu\tau}+\phi_2)} \\ 
|m_{\nu_e\nu_\tau}|e^{-i(\phi_{e\tau}+\phi_2-\phi_1)} && |m_{\nu_\mu\nu_\tau}|e^{-i(\phi_{\mu\tau}+\phi_2)} && |m_{\nu_\tau\nu_\tau}| 
\end{pmatrix*}; 
\end{equation}
\end{widetext}

where $M_\nu^\prime=M_\phi^\dagger M_\nu M_\phi$. Three cases are present in this 
expression:\\

{\bf Case A}: $\phi_1=\phi_{e\mu}$ and 
$\phi_2=\phi_1-\phi_{e\tau}=\phi_{e\mu}-\phi_{e\tau}$. Producing
\begin{equation}\label{mngepA}
M_\nu^\prime=\begin{pmatrix*}[l]
|m_{\nu_e\nu_e}| && |m_{\nu_e\nu_\mu}| && |m_{\nu_e\nu_\tau}| \\ 
|m_{\nu_e\nu_\mu}| && |m_{\nu_\mu\nu_\mu}| && |m_{\nu_\mu\nu_\tau}|e^{i\psi} \\
|m_{\nu_e\nu_\tau}| && |m_{\nu_\mu\nu_\tau}|e^{-i\psi} && |m_{\nu_\tau\nu_\tau}| 
\end{pmatrix*}; 
\end{equation}
with $\psi=\phi_{\mu\tau}+\phi_2=\phi_{\mu\tau}+\phi_{e\mu}-\phi_{e\tau}$.\\

{\bf Case B}: $\phi_1=\phi_{e\mu}$ and 
$\phi_2=-\phi_{\mu\tau}$. Producing
\begin{equation}\label{mngepB}
M_\nu^\prime= \begin{pmatrix*}[l] |m_{\nu_e\nu_e}| && |m_{\nu_e\nu_\mu}| && |m_{\nu_e\nu_\tau}|e^{-i\psi} \\ 
|m_{\nu_e\nu_\mu}| && |m_{\nu_\mu\nu_\mu}| && |m_{\nu_\mu\nu_\tau}| \\ 
|m_{\nu_e\nu_\tau}|e^{i\psi} && |m_{\nu_\mu\nu_\tau}| && |m_{\nu_\tau\nu_\tau}| 
\end{pmatrix*}.
\end{equation}\\

{\bf Case C}: $\phi_2=-\phi_{\mu\tau}$ and  
$\phi_1=\phi_2+\phi_{e\tau}=\phi_{e\tau}-\phi_{\mu\tau}$. Producing
\begin{equation}\label{mngepC}
M_\nu^\prime=\begin{pmatrix*}[l] |m_{\nu_e\nu_e}| && |m_{\nu_e\nu_\mu}|e^{i\psi} && |m_{\nu_e\nu_\tau}| \\ 
|m_{\nu_e\nu_\mu}|e^{-i\psi} && |m_{\nu_\mu\nu_\mu}| && |m_{\nu_\mu\nu_\tau}| \\ 
|m_{\nu_e\nu_\tau}| && |m_{\nu_\mu\nu_\tau}| && |m_{\nu_\tau\nu_\tau}| 
\end{pmatrix*}; 
\end{equation}

From the former, we can conclude that using a WBT 
we can get rid of two unwanted phases, ending up with a single phase responsible 
for the possible CP violation phenomena present in the PMNS mixing matrix.

We find that in matrix \ref{mngepA} (or 
equivalently in \ref{mngepB} or \ref{mngepC}), the final number 
of parameters are 
six real numbers and one phase $(\psi)$, just enough to accommodate the neutrino oscillation parameters. Further texture zeros will give relationships between neutrino masses and mixing parameters.

In this way, one texture zero would allow us to write one of the mixing angles 
$\theta_{ij}$ as a function of the neutrino masses; meanwhile, two texture zeros allow us to 
write two mixing angles as a function of the three neutrino masses. Three or more texture zeros are too restrictive and not in agreement with the experiments.

The most important consequence of the former analysis is that, since the phases $\phi_1$ and $\phi_2$ are arbitrary, the remaining phase $\psi$ obtained after the weak basis transformation can be placed in any entry of the neutrino mass matrix, according to equations 
\ref{mngepA}-\ref{mngepC}. In particular, if we impose an off-diagonal vanishing element, we can rotate the remaining phase to that entry, and the mass matrix becomes real, implying CP conservation.

\bibliographystyle{apsrev4-2}
\bibliography{main}

\end{document}